\documentclass[pdflatex,sn-mathphys-num]{sn-jnl}
\usepackage{graphicx}%
\usepackage{multirow}%
\usepackage{amsmath,amssymb,amsfonts}%
\usepackage{amsthm}%
\usepackage[title]{appendix}%
\usepackage{xcolor}%
\usepackage{textcomp}%
\usepackage{manyfoot}%
\usepackage{booktabs}%
\usepackage{algorithm}%
\usepackage{algorithmicx}%
\usepackage{algpseudocode}%
\usepackage{listings}%
\usepackage{bm}

\theoremstyle{thmstyleone}%
\theoremstyle{thmstyletwo}%

\theoremstyle{thmstylethree}%

\graphicspath{{Figure/}}

   \usepackage[caption=false,font=footnotesize]{subfig}

\usepackage{mathrsfs}%

\begin{document}
\title[Passive Indoor Localization with WiFi Fingerprints]{Passive Indoor Localization with WiFi Fingerprints}
\author[1]{\fnm{Minh Tu} \sur{Hoang}}\email{hoangminhtu90@gmail.com}
\author[1]{\fnm{Brosnan} \sur{Yuen}}\email{brosnanyuen@gmail.com}
\author[1]{\fnm{Kai} \sur{Ren}}\email{ryankey99@hotmail.com}
\author[1]{\fnm{Ahmed} \sur{Elmoogy}}\email{ahm.magdy90@gmail.com}
\author*[1]{\fnm{Xiaodai} \sur{Dong}}\email{xdong@uvic.ca}
\author*[1]{\fnm{Tao} \sur{Lu}}\email{taolu@uvic.ca}
\author[1]{\fnm{Hung Le} \sur{Nguyen}}\email{nlhung279@uvic.ca}
\author[2]{\fnm{Robert} \sur{Westendorp}}\email{rwestendorp@fortinet.com}
\author[2]{\fnm{Kishore Reddy} \sur{Tarimala}}\email{kreddy@fortinet.com}
\affil*[1]{\orgdiv{Department of Electrical and Computer Engineering}, \orgname{University of Victoria}, \orgaddress{\street{3800 Finnerty Rd.}, \city{Victoria}, \state{BC}, \postcode{V8P 5C2}, \country{Canada}}}
\affil[2]{\orgname{Fortinet Canada Inc.}, \city{Burnaby}, \state{BC},  \postcode{V5C 6C6}, \country{Canada}}

\abstract{
This paper proposes passive WiFi indoor localization. Instead of using WiFi signals received by mobile devices as fingerprints, we use signals received by routers to locate the mobile carrier. Consequently,  software installation on the mobile device is not required. To resolve the data insufficiency problem, flow control signals such as request to send (RTS) and clear to send (CTS) are utilized. In our model, received signal strength indicator (RSSI) and channel state information (CSI) are used as fingerprints for several algorithms, including deterministic, probabilistic and neural networks localization algorithms. We further investigated localization algorithms performance through extensive on-site experiments with various models of phones at hundreds of testing locations. We demonstrate that our passive scheme achieves an average localization error of  $\sim$0.8~m when the phone is actively transmitting data frames and $\sim$1.5~m when it is not transmitting data frames.}

\keywords{Passive indoor localization, channel state information (CSI), received signal strength indicator (RSSI), WiFi indoor localization. }
\maketitle

\section{Introduction} \label{sec:intro} 
Indoor localization using WiFi fingerprints enables a wide variety of applications, including user tracking, location-based advertising, and virtual reality gaming, etc.~\cite{He2016, Zafari2019}. In general, WiFi localization systems can be categorized into  active and passive schemes~\cite{Xu2016, Deak2012}. The active localization scheme, being the mainstream approach, collects WiFi signals from a mobile device to infer the user's positions~\cite{Minh2018,Minh2020}. It requires a dedicated software to be installed on mobile devices for WiFi scanning and logging~\cite{Duan2020}. In contrast, passive localization is fully implemented on WiFi access points (APs) and does not require any software to be installed on mobile devices~\cite{Deak2012, Chen2016}. 

WiFi passive indoor localization has attracted widespread interests in recent years \cite{Cui_2015}. In the past, device-free passive localization that does not require user to carry any mobile devices have been studied. By extracting channel state information (CSI) fluctuation from multiple APs induced by human movements, Shi \textit{et al.}~\cite{Shi2018} proposed a probabilistic scheme to track a user's location. Using received signal strength indicator (RSSI) instead of CSI, Liu \textit{et al.}~\cite{Chen2016}  locates the position by matching the RSSI testing values to their distribution properties stored in the database. Although convenient for users, the estimation error of device-free passive localization can be as high as 3.5 m at a density of one sampling point per meter. Meanwhile, such algorithms can only locate a few users~\cite{Luo2017}. For higher resolution, larger localization area and more users, passive localization with mobile devices is preferred. Since the data rates at different locations are different, Duan \textit{et al.}~\cite{Duan2019} exploits data rate information of all available APs as the fingerprints for localization. In order to improve the resolution, their subsequent work~\cite{Duan2020} includes other fingerprints such as packet delivery ratio (PDR) for localization, which improves the average accuracy to 2.5~m.

This paper focuses on WiFi passive indoor localization with the following main contributions.
\begin{itemize}
\item[1] A comprehensive practical passive indoor localization study is conducted for the first time, with detailed analysis for the cases when the phone is  active or is inactive.  We use RTS/CTS to increase the packet rate for inactive phones, of which increases localization accuracy. Meanwhile, we incorporate CSI as part of the fingerprints in order to boost the localization accuracy for  active phones.

\item[2] An efficient mitigation scheme for both MAC randomization and device heterogeneity problems is demonstrated.

\item[3] Appropriate localization algorithms are selected for different phone states. For inactive phones, CSI is not available, and thus we have to solely depend on RSSI. However, RTS/CTS method allows us to get more packets for RSSI algorithms such as SRL-KNN~\cite{Minh2018}, P-MIMO LSTM~\cite{Minh2020}, and SSP~\cite{Minh2019}. For active phones, CSI and RSSI are both available. A method to combine both fingerprints to enhance the location accuracy is proposed.

\item[4] The proposed algorithms are tested in an extensive autonomous experiment with a robot, thousands of RPs, several testing trajectories and a variety of phone types such as Samsung Galaxy S6, HTC One X, iPhone X and LG Nexus 5.
\end{itemize}   

\section{Related work} \label{sec:related_work} 
Table~\ref{table:ExpCompare} summarizes  the current literature on WiFi indoor localization and their respective experiments. The number of APs, RPs and testing points vary among different experiments, and the grid size is defined as the average distance between two adjacent RPs. Among all, the largest number of APs is 15~\cite{Fang2008}. The largest number of RPs and testing points are 207 and 50 respectively~\cite{Battiti2002}, whereas most experiments incorporate fewer than 100 RPs with a maximum number of 6 APs.
\begin{sidewaystable}
\caption{Current Literature on Indoor Localization} \label{table:ExpCompare} 

\begin{tabular}{|l| c| c| c| c| c| c| c| c|} 
\hline           
\textbf{Method} & \textbf{Fingerprint}  & \textbf{Class} & \textbf{Access point (AP)} & \textbf{Reference point (RP)} & \textbf{Testing Point} & \textbf{Grid Size} & \textbf{Accuracy}\\ 
\hline
FILA~\cite{Wu2013} & CSI  & Active &  1-3 &  28 &  -  & - & 0.4 m to 1 m \\
DeepFi~\cite{Wang2017} & CSI & Active &  1 &  50 &  30  & - & 0.94 $\pm$ 0.56 m\\
ConFi~\cite{Haochen2017} & CSI & Active &  1 &  64 &  10   & - & 1.3 $\pm$ 0.9 m\\
PhaseFi~\cite{Wang2016a} & CSI & Active &  1 &  38 &  12  & - &1.0 $\pm$ 0.4 m\\
CiFi~\cite{Wang2018} & CSI & Active &  1 &  15 &  15  & - &1.7 $\pm$ 1.2 m\\
BiLoc~\cite{Wang2017b} & CSI & Active & 1 & 25 & 25 & 1.8 m & 1.5 $\pm$ 0.8 m\\
MLP~\cite{Battiti2002} & RSSI & Active &  6 &  207 &  50  & 1.7 m & 2.8 $\pm$ 0.1 m\\
DANN~\cite{Fang2008} & RSSI & Active &  15 &  45 &  46  & 2 m & 2.2 $\pm$ 2.0 m\\
RELM~\cite{Lu2016} & RSSI & Active &  8 &  30 &  10  & 3.5 m & 3.7 $\pm$ 3.4 m\\
MLNN~\cite{Dai2016} & RSSI & Active &  9 &  20 &  20  & 1.5 m & 1.1 $\pm$ 1.2 m\\
DR Fingerprinting~\cite{Duan2019} & DR & Passive & 4 & 18 & 47 & 5.9 m & 3.4 m\\
PDR Fingerprinting~\cite{Duan2020} & PDR & Passive & 8 & 60 & - & 3.3 m & 2.6 m\\
\hline         
\end{tabular} 
\end{sidewaystable}

\begin{figure}[!t]
\centering
\includegraphics[width=0.8\textwidth]{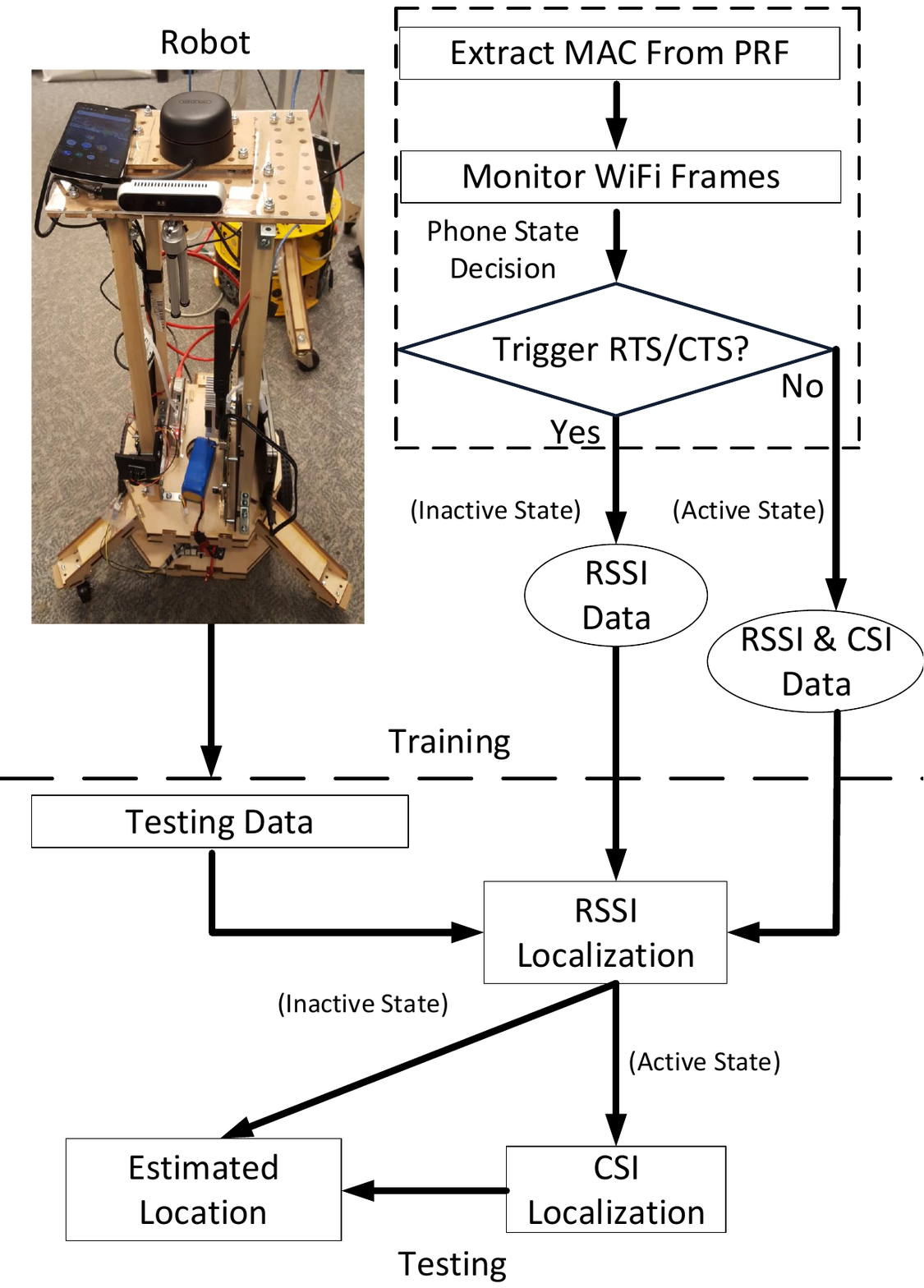}
\caption{Proposed Localization Process}
\label{fig:process1}
\end{figure}

\begin{figure*}[!t]
\centering
\includegraphics[width=\textwidth]{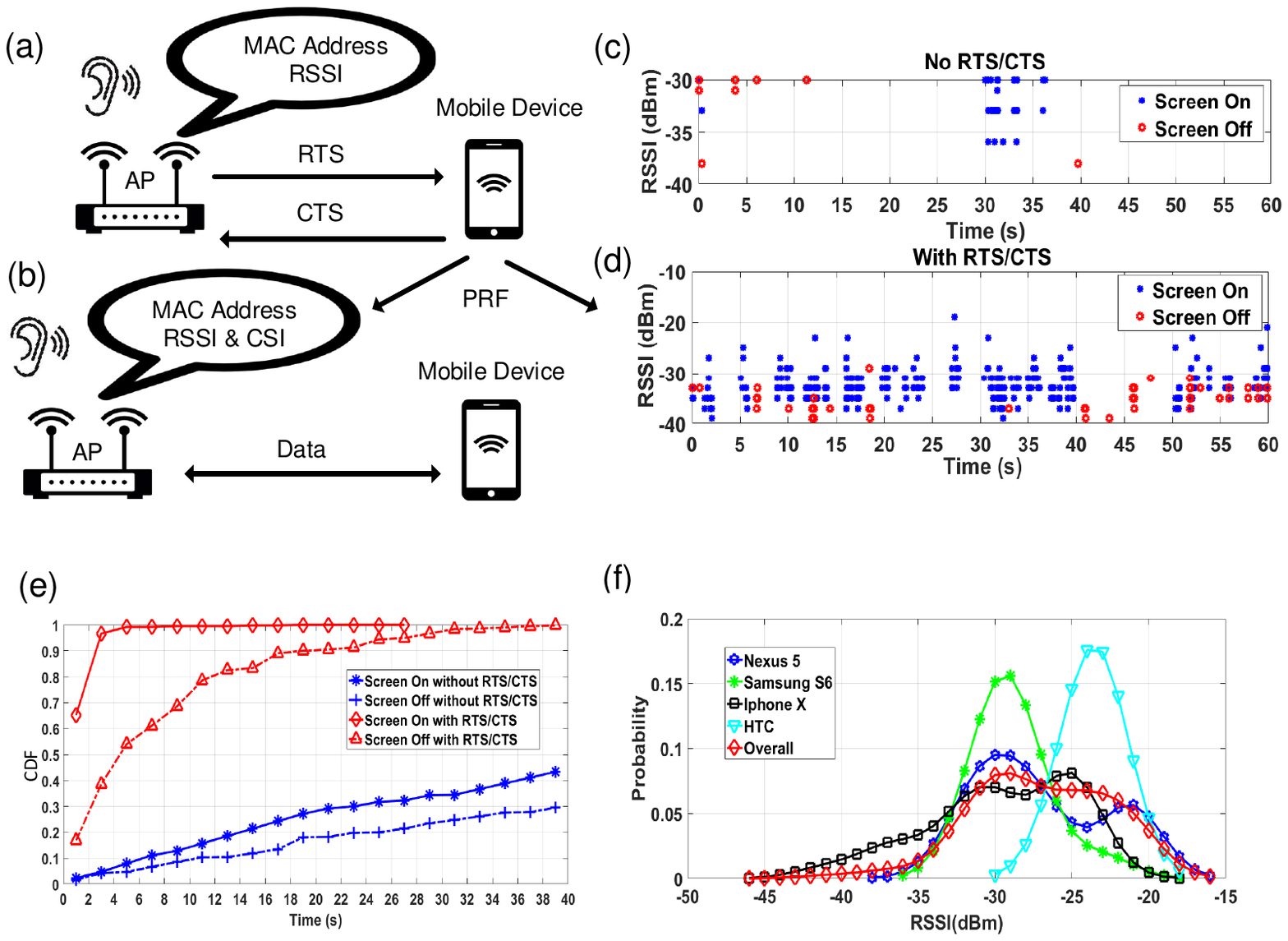}
\caption{Data transmission scheme when: (a) the phone is inactive and the RTS/CTS mechanism is executed, (b)  the phone is actively exchanging data frames with the AP, (c) the phone's transmitted packets are plotted over time while the phone is inactive, (d) the phone's transmitted packets are plotted over time while the phone is responding to RTS/CTS packets and is not connected to anything, (e) the cumulative distribution functions of the Samsung Galaxy S6's transmitted packets are plotted over time, and (f) PDF of RSSI for various devices: Nexus 5 (blue star), Samsung S6 (green square), iPhone X (black triangle), and the overall PDF (red circle) at a fixed location.}
\label{fig:process}
\end{figure*}

\subsection{Indoor Localization Algorithms}
Fingerprinting based WiFi localization algorithms can be classified into deterministic and probabilistic approaches~\cite{He2016}. The former uses a similarity metric to compare the measured signal and the fingerprint data in the database for location estimation. Some typical examples of this approach are artificial neural network (ANN)~\cite{Wang2017,Minh2019}, support vector machine (SVM)~\cite{Brunato2005,Shi2015} and K nearest neighbours (KNN)~\cite{Minh2018,YaqinXie2016}, all of which require the collection of the fingerprints in the training phase to be compared with the measured signal in the testing phase. The simplest deterministic approach is KNN that determines the user location by calculating and ranking the fingerprint distance measured at the unknown point and the reference points (RPs) in the database~\cite{YaqinXie2016, DongLi2016, Minh2018}. Moreover, SVM~\cite{Brunato2005} provides a direct mapping from RSSI values collected at the mobile devices to the estimated locations through nonlinear regression by supervised classification technique~\cite{C.Gentile2013}. Despite their low complexity, the accuracy of these methods are unstable due to the wide fluctuation of WiFi signals~\cite{YaqinXie2016,DongLi2016}. In contrast, ANN~\cite{Haochen2017, Minh2019, Ayya_2020} estimates location non-linearly from the input by a chosen activation function and adjustable weightings. In indoor environments, because the mapping between the WiFi fingerprints and the user's locations is nonlinear, it is difficult to formulate a closed-form solution~\cite{C.Gentile2013}. ANN provides more suitable and reliable solutions for its ability to approximate high dimension and highly nonlinear models~\cite{Minh2019,Minh2020_2}. On the other hand, the probabilistic method is based on statistical inference between the target signal readings and stored fingerprints using Bayes rule~\cite{Minh2020}. In order to determine the location, probabilistic methods require the probability density function (PDF) of fingerprints. Early research assume the RSSI PDF follows empirical parametric distributions such as  single-peak~\cite{Kaemarungsi2004} or double-peak Gaussian~\cite{L.Chen2013}, log normal distribution~\cite{Honkavirta2009}, etc. To achieve better performance by eliminating the assumption on the RSSI PDF, Kernel methods~\cite{C.Figuera2009, Minh2020} employ component smoothing functions for each data value to produce a smooth and continuous probability curve and avoid any zero count bins.


\subsection{Device Heterogeneity} \label{subsec:heter} 
Different mobile devices have different power transmission circuitry, leading to different signal strengths at the same location~\cite{He2016}.  This causes device heterogeneity and is the primary challenge of indoor localization~\cite{Baun2010}. If the WiFi device used in the training phase is different from the actual device used in the testing phase, the localization accuracy could be significantly reduced~\cite{Duan2020}. To mitigate this problem, one may use special fingerprints such as data rate (DR)~\cite{Duan2019}, RSS ratio~\cite{Cheng2013}, RSSI difference~\cite{Hossain2011} to avoid heterogeneity. However, due to power control on the mobile devices, the number of frames or power sent from a mobile device may vary significantly, which makes all of those fingerprints less reliable. Alternatively, M. Kjærgaard~\cite{Baun2010} proposes to calibrate RSSI from two different devices by two different signal strengths using a linear model. However, the linear model neglects the non-linear characteristics of WiFi signals in fading environments and causes estimation error. To enhance the performance, J. Park et al.~\cite{Park2011} uses the RSSI distribution rather than each single RSSI value. In this case, kernel density function is used to reduce the difference between two RSSI distributions from two different devices. The test is conducted with 3 laptops and 2 phones in 18 RPs with the average accuracy being around 2~m.

To summarize, all the proposed WiFi localization approaches have the following limitations:
\begin{itemize}
\item[1] Active localization approaches require dedicated software installed on mobile devices to perform WiFi scanning and logging data for the localization process, which is inconvenient for users. On the other hand, the passive localization is more practical but still not fully characterized with comprehensive tests.

\item[2] The existing passive localization experiments are limited to insufficient numbers of RPs and testing points per unit testing area because the measurements on the RPs are conducted manually. As the localization accuracy is determined by the density of the RPs in the target area, the performances of the reported algorithms are limited. Further, due to the limited number of measurements conducted manually, the previous works select fixed testing points or exactly on the same spots of the RPs and treat the localization as a classification model, which further departs the model generalization from practical cases.

\item[3] The analysis of different states of the phones are insufficiently investigated. In practical scenarios, phones behave differently due to usage data requirements. For the purposes of this paper, a phone is active when it is transmitting data frames. A phone is inactive when it is not transmitting data frames. Consequently, existing passive indoor localization schemes that uses data rate~\cite{Duan2019} or packet delivery ratio~\cite{Duan2020} as fingerprints may not have sufficient data for localization when the phone is inactive. 
\end{itemize}  

On the other hand, besides WiFi, some researchers also use Bluetooth to locate the indoor position~\cite{Bai2020,Nikodem2021} and especially the passive localization in BluePIL~\cite{Rodrigues2021}. The advantage of using Bluetooth technology is that it consumes lower energy and is not impaired by MAC randomization issues in WiFi localization. However, the coverage range of Bluetooth is much shorter than WiFi router. Therefore, in order to have the same coverage area and accuracy, many more Bluetooth beacons are required. Furthermore, WiFi routers are much more populated in most practical environments compared to the Bluetooth counter part, which makes WiFi be a better choice. The following sections will propose detailed solutions to the problems listed above for both active and inactive phones. 

\section{Proposed Passive Localization System} \label{sec:proposed} 
As shown in Fig.~\ref{fig:process1}, the proposed localization is divided into two phases: a training phase and a testing phase. In the training phase, fingerprints from all available $P$ APs are collected at each of the $M$ predefined RP locations and stored in a database.  Here we define a fingerprint vector at RP $i$ to be $\bm{F}(\bm{l}_i)=\{F_{1}(\bm{l}_i), F_{2}(\bm{l}_i),...,F_{N}(\bm{l}_i)\}$, where $\bm{l}_{i}{\equiv}(x_{i},y_{i})$ is the physical location of RP $i$, $N$ is the number of features of the fingerprints and $F_{j}(\bm{l}_i),\, 1 \leq j \leq N,$ is the $j$-th feature at RP $i$. In the testing phase, our localization algorithms determine the position of a user carrying a mobile device with the only assumption that the device WiFi is on.

Both training and testing phases start with the phone state determination steps. When a mobile phone is within the target area, all available $P$ APs detect the probe request frames (PRF) sending from the phone. According to the WiFi standard IEEE 802.11 \cite{IEEE_80211}, a mobile device periodically broadcasts PRF to discover nearby networks~\cite{Martin2017}. Therefore, each AP will receive at least one PRF when the phone WiFi is on. After PRF is successfully detected, the media access control (MAC) address of the device can be extracted. Subsequently, the AP continues to monitor all the frames from that MAC address to determine whether the number of received WiFi frames are sufficient for localization. Evidently,  higher localization accuracy requires higher sampling rate. In practical scenarios, the user location will be updated at a consecutive sampling time interval $\Delta{t}$ with the requirement that at least one WiFi frame is received in that interval. In general, this requirement can be easily met when the phone is active. However, the number of frames per $\Delta{t}$ is significantly reduced when the phone is inactive. In order to achieve consistent accuracy regardless of the phone states, we propose separate localization solutions to inactive phones as shown in Fig.~\ref{fig:process}(a) and active phones shown in Fig.~\ref{fig:process}(b).

\subsection{Inactive Phone Localization}~\label{sec:ideal}
In this case, the mobile phone does not transmit data frames. Moreover, the packet rate of other WiFi frames are severely limited to save the battery power. If the phone screen is locked or turned off, the number of WiFi frames will be even lower. Fig.~\ref{fig:process}(c) illustrates the packet transmission timeline of an inactive Samsung Galaxy S6 phone. WiFi frames are monitored by Alfa AC1200 router in a 2.4 GHz channel. As shown, within 60~seconds, when the screen is on, the phone sends frames infrequently with a time gap as large as 20 seconds (blue dots). Further, when the screen is off, the time gap between frames increases to more than 40 seconds (red dots). Fig.~\ref{fig:process}(d) shows the cumulative distribution function (CDF) of the time interval between an Alfa AC1200 router receiving two consecutive packets from Samsung Galaxy S6. The experiment was conducted once a day with the phone at various battery levels. As shown, the probability that a second WiFi data frame transmission within 30 seconds is less than 30$\%$ (blue line), which is not enough for continuous localization. Therefore, we need to apply a solution to increase the data transmission.           

Here we propose to apply RTS/CTS process to overcome this challenge. RTS/CTS is a handshaking mechanism to reserve the channel for a specific duration before the actual data transfer starts~\cite{Sawwashere2014}. In Fig.~\ref{fig:process}(a), after obtaining the MAC address of the target phone from PRF, the AP will send $K$ RTS frames to that MAC address every $\Delta{t}$. According to the IEEE 802.11 standard, when an AP transmits an RTS frame to a target, the target should immediately reply a CTS frame to the AP even if it is not associated with the AP~\cite{Duan2020}. Therefore, the phone will be forced to respond with several CTS frames back to the corresponding AP. Fig.~\ref{fig:process}(d) illustrates the timeline of sending frames from a Samsung Galaxy S6 phone in the same 60~s period with an RTS frame sending from AP every $200$ ms. Apparently, the number of packets carrying CTS frames sent from the phone increases significantly comparing with the case without RTS/CTS process. In the case when the screen is on, Fig.~\ref{fig:process}(e) shows that in 95$\%$ probability that the router receives another WiFi frame within 4~seconds after the first frame was sent, which enables continuous localization. 

As  all available APs receives the packets carrying CTS, they can identify the phone with the assist of a MAC filter and extract the RSSI information. Therefore, through this approach, we can collect sufficient fingerprints and apply localization algorithms described in Subsection~\ref{sec:algorithm}.     
    
\subsection{Active Phone Localization} \label{sec:trans}
When the phone is active, it transmits data frames together with ACK, CTS, and probe request frames. Therefore, use of RTS/CTS is unnecessary. With the long preamble in a data frame~\cite{Matthias2018}, both RSSI and CSI can be extracted from the APs at a satisfactory sampling rate. In contrast to inactive phones where only RSSI is available from CTS, the localization accuracy can be even higher by combining both RSSI and CSI as fingerprints. The algorithm detailed with the combination of RSSI, CSI fingerprints will be presented in Subsection~\ref{sec:algorithm} and Subsection~\ref{sec:trans_result}.

\subsection{Localization Algorithms} \label{sec:algorithm} 
 For high accuracy while mitigating the effects of device heterogeneity mentioned in Subsection~\ref{subsec:heter}, we propose different sets of localization algorithms for active phones and inactive phones. 
\subsubsection{Inactive Phone Localization}
In this scenario, the only available fingerprint is RSSI, the values of which vary among different mobile devices. Fig. \ref{fig:process}(f) shows the RSSI PDF of Nexus 5, Samsung Galaxy S6, HTC One X and iPhone X phones placed at the same location within a period of one minute. The RSSI range of each phone received by a nearby AP varies significantly. The RSSI from iPhone X ranges from -45 dBm to -18 dBm, while Nexus 5 between -38 dBm and -16 dBm, HTC between -30 dBm and -18 dBm, and Samsung S6 from -36 dBm to -18 dBm. Since RSSI from each phone has different distributions, using the RSSI data from a particular phone can not train a machine learning model that suits all other phones. Therefore, RSSI data from as many devices as possible should all be used to train a generic machine learning model.  We use 4 different phones including Samsung Galaxy S6, HTC One X, iPhone X and LG Nexus 5 (Subsection \ref{sec:experiment}) for training.  

Probabilistic methods with Kernel density estimator described in~\cite{C.Figuera2009,Park2011} can represent RSSI features over several devices well. The Kernel density function can estimate RSSI PDF of AP $k$, $P(F_{k}(\bm{l}_{i})|\bm{l}_{i})$, for each location $\bm{l}_{i}$ which can includes multiple RSSI values from multiple phones.
\begin{equation}
P(F_{k}(\bm{l}_{i})|\bm{l}_{i}) = \frac{1}{nh} \sum_{x=1}^{n} K\left(\frac{F_{k}(\bm{l}_{i}) - F^{x}_{k}(\bm{l}_{i})}{h}\right)  
\end{equation}
$n$ is the number of RSSI samples in location $\bm{l}_{i}$. $F^{x}_{k}(\bm{l}_{i})$ is RSSI fingerprint sample collected in location $\bm{l}_{i}$. $F_{k}(\bm{l}_{i})$ is the random variable of the RSSI at $\bm{l}_{i}$. K(·) is the Kernel smoothing function, and $h$ is the bandwidth. Details and analysis for choosing different kind of Kernel smoothing functions including Gaussian, local linear, Kernel average, etc., and bandwidth can be found in~\cite{C.Figuera2009,Park2011,Minh2020}. We employ this technique to estimate RSSI PDF along with semi-sequential SSP model~\cite{Minh2020} to achieve high localization accuracy. An overall PDF $P(F_{k}(\bm{l}_{curr})|\bm{l}_{i})$ in the Eq.~\eqref{prob5} below that takes into account RSSIs from all phones is estimated using the Kernel density estimator and illustrated as the red circle line in Fig.~\ref{fig:process}(f).  The PDF will then be used in SSP model for localization~\cite{Minh2020} 
\begin{equation}  \label{prob5}
P(\bm{l}_{curr}\approx\bm{l}_{i}|\bm{F}(\bm{l}_{curr}),\bm{l}^{\prime}_{pre})\propto \\
\prod_{k=1}^{N} P(F_{k}(\bm{l}_{curr})|\bm{l}_{i})P(\bm{l}_{i}|\bm{l}^{\prime}_{pre}),
\end{equation}
where $\bm{F}(\bm{l}_{curr})$ is the fingerprint vector of the unknown current location $\bm{l}_{curr}$, while $\bm{l}^{\prime}_{pre}(x_{pre}^\prime,y_{pre}^\prime)$ is the predicted previous location. $ P(\bm{l}_{i}|\bm{l}^{\prime}_{pre})$ is a soft range SSP window that may have the shape of Gaussian, Hann, or Tukey. The likelihood function $P(F_{k}(\bm{l}_{curr})|\bm{l}_{i})$ describes the probability of the $k$-th fingerprint feature at location $\bm{l}_i$.   

In addition, recurrent neural network (RNN) exploits the additional information in time domain and is effective in localization~\cite{Minh2019}. Instead of relying on one instant RSSI scan, RNN correlates the previously predicted positions within the trajectory to predict the subsequent position.  Therefore, it reduces the device-dependency on RSSI localization. Among all RNN topologies, P-MIMO LSTM~\cite{Minh2019} provides the best accuracy with the robustness in multiple databases and environments. Therefore, in this paper, we choose to implement P-MIMO LSTM.   

Due to substantial RSSI fluctuations in dynamically changing environments such as human blocking and movements, we applies weighted filter in P-MIMO LSTM to filter out the outliers~\cite{Minh2019}. In addition, SSP model relies on the statistical feature of a multiple RSSIs instead of one, which mitigates the prediction inaccuracy from the RSSI fluctuations~\cite{Minh2020}.  

\subsubsection{Active Phone Localization}
When the phone is active, both RSSI and CSI are available and their transmission rate is sufficient for accurate localization. The device heterogeneity problem arising from RSSI of different  phones at the same location being significantly different, still impairs the localization accuracy, which however, can be mitigated partially with the additional CSI information incorporated. In our implementation, CSI amplitudes are grouped into a CSI-image-like~\cite{Haochen2017, Minh2020_2} time-frequency matrix to form fingerprints at each location. Meanwhile, the raw CSI phase is pre-processed extracting the phase difference between sub-carriers~\cite{Wang2018}, is needed before being used as an additional feature~\cite{Haochen2017} to reduce the inaccuracy from noises and random fading~\cite{Wang2017}.    

\begin{figure}[!t]
     \centering
\subfloat[\label{fig:CSI_Image}]
{\includegraphics[width=1\textwidth]{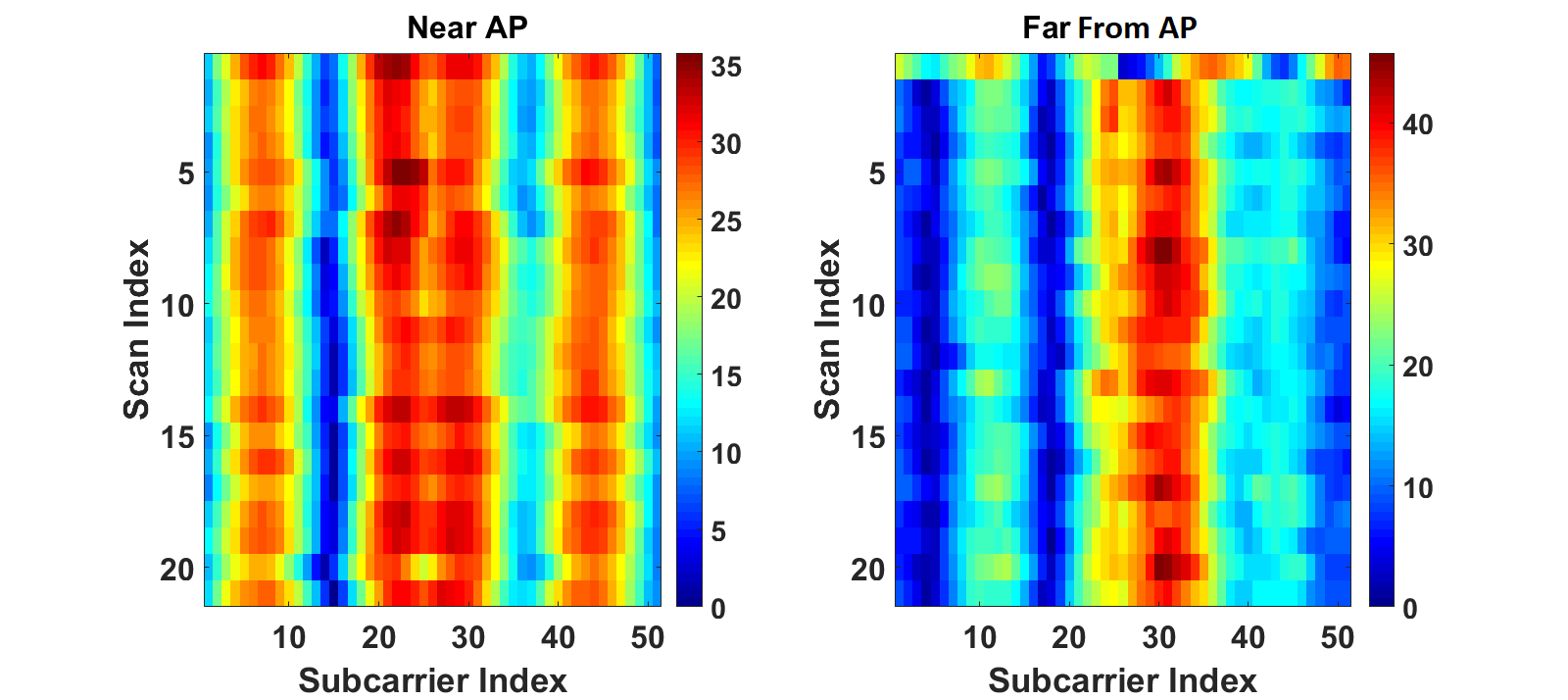}} \quad
\subfloat[\label{fig:CNN_phase}]
{\includegraphics[width=1\textwidth]{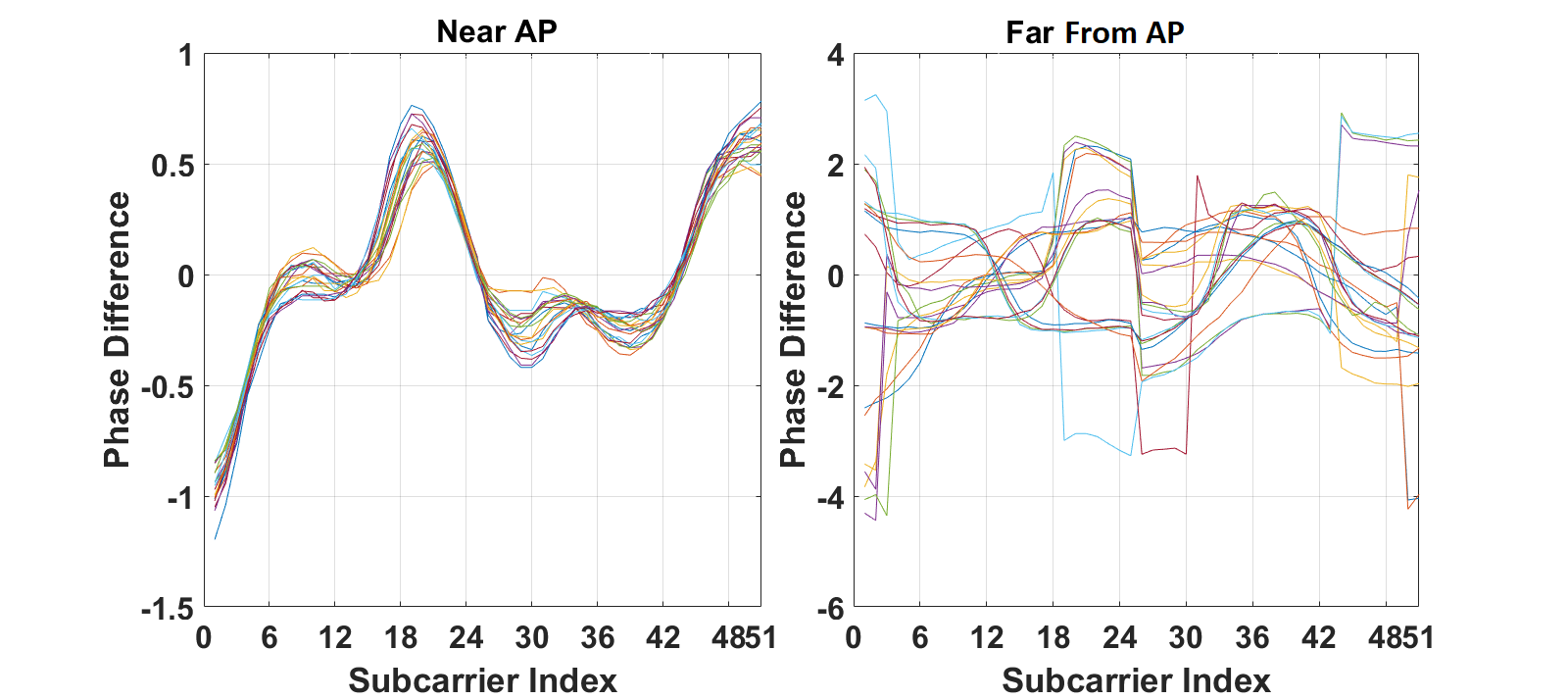}}
\caption{CSI from a Samsung Galaxy S6 phone at 2 fixed locations (a) CSI amplitude images. (b) CSI phase difference.}
     \label{fig:CSI_example}
\end{figure}
To illustrate, ESP32~\cite{esp32}, a low-cost, low-power system-on-a-chip microcontroller with integrated WiFi, is attached to each router to collect CSI at a 20~MHz bandwidth and 2.4~GHz frequency.  In total, 51 sub-carriers are extracted from 20 scans of received packets using a Samsung Galaxy S6 phone at 2 representative locations: one location is within $1~$m distance to the AP and with line of sight (LOS); the other location is more than $10~$m away from the AP and does not have LOS (NLOS). Fig.~\ref{fig:CSI_example}(a) presents the CSI amplitude image constructed by the method in~\cite{Minh2020_2}, while Fig.~\ref{fig:CSI_example}(b) shows the corresponding phase difference information following the method in CIFI~\cite{Wang2018}.  Clearly, CSI at the near AP location is more stable than the one far away. Further, CSI phase from the nearby location is similar in each scan, while that from far away fluctuates significantly. The reason is that CSI phase is sensitive to environments especially the interference from human movements~\cite{Minh2020_2}. Therefore, the larger the distance is between the phone and AP, the more probable that the environment disturbances will reduce the localization accuracy~\cite{Minh2020_2}.

To incorporate both RSSI and CSI, our proposed method for the active phone localization (Fig.~\ref{fig:process1}) takes a two-step approach. We first adopt RSSI localization using SSP or P-MIMO LSTM to pre-determine $K$ possible nearest neighbour locations. In a second step, the CSI information is utilized to select the most likely position among those $K$ candidates. Since CSI amplitude and phase are stable only if the signal is from the location near APs, we only select $L$ APs that provide the strongest signals for localization. In testing phase, the CSI amplitude and phase of a testing point is correlated with the CSI information of the $K$ neighbour locations using Pearson coefficient~\cite{Minh2018} and the one that provides the largest coefficient is predicted to be the location.         


\section{Database And Experiments} \label{sec:experiment}
\begin{figure*}[!t]
\centering
\includegraphics[width=\textwidth]{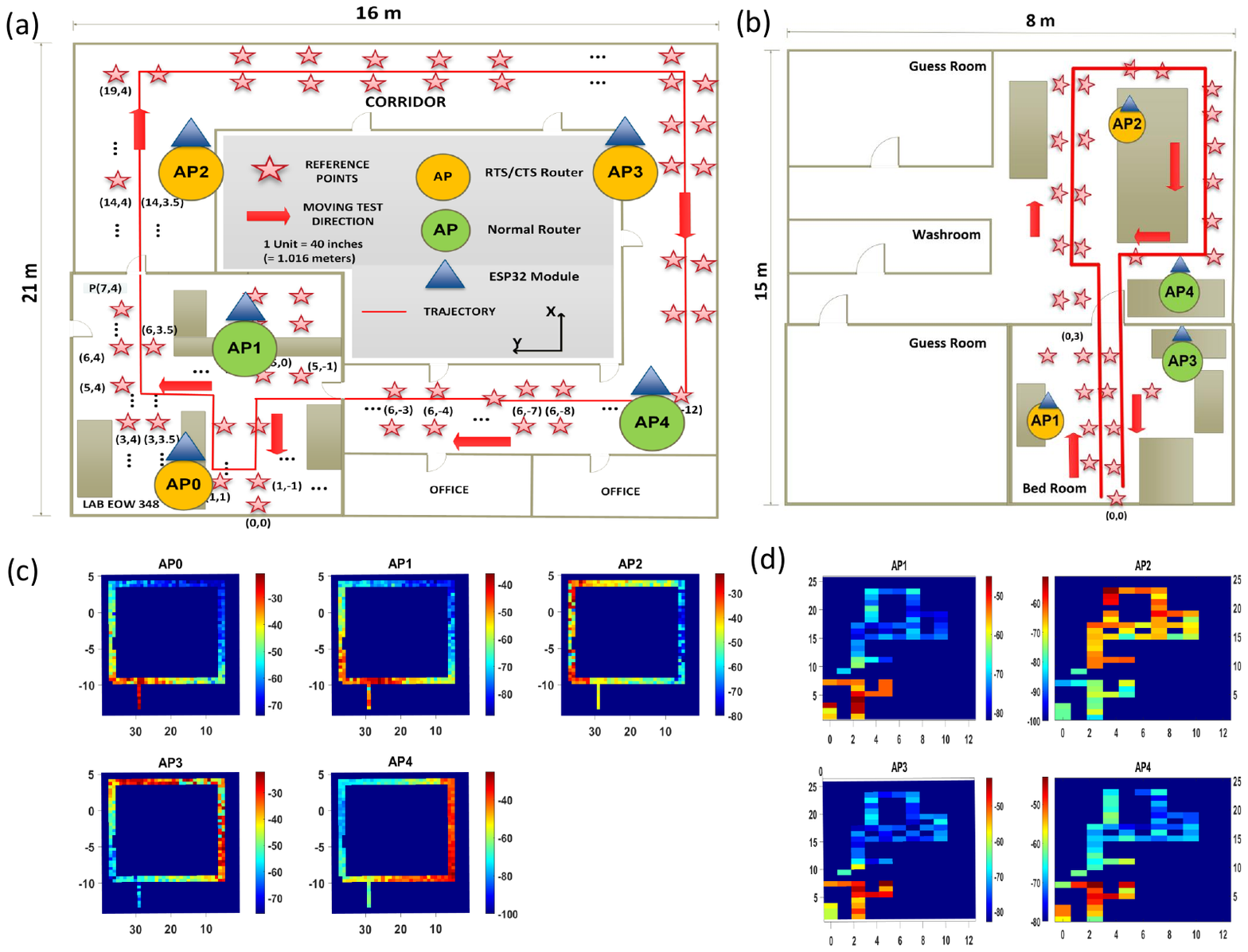}
\caption{(a) Office experiment floor map. (b) Home experiment floor map. (c) RSSI heat map of office experiment. (d) RSSI heat map of home experiment.}
\label{fig:floor_map}
\end{figure*}

All experiments have been carried out in 2 different settings: office and home. The experiments in office setting are conducted on the third floor of Engineering Office Wing (EOW), University of Victoria, BC, Canada with the floor plan showing in Fig.~\ref{fig:floor_map}(a). The dimension of the area is 21~m$~\times$~16~m. On the other hand, the home setting related experiments  are conducted in the basement of a two-storey single family house in Victoria, BC, Canada with the area being 15~m~$\times$~8~m as shown in Fig.~\ref{fig:floor_map}(b). The fingerprints for both training and testing are collected using an autonomous driving robot. The 3-wheel robot as shown in Fig.~\ref{fig:process1} has multiple sensors including a wheel odometer, an inertial measurement unit (IMU), a LIDAR, 3 sonar sensors and a colour and depth (RGB-D) camera. It can navigate to a target location within an accuracy of $\mathrm{0.21{\pm}0.02~m}$.

During the experiments, the robot carried each one of the four the mobile phones in turn to collect the WiFi fingerprints. In the training phase, the robot stays at every RP for 2 minutes to collect data. The data are repeatedly collected in different days and at different time slots to build a comprehensive database that covers both phone states. In the testing phase, the robot navigates along a pre-defined route showing as the red traces in Fig.~\ref{fig:floor_map}(a) and (b) at speeds randomly generated within 0.6-4.0~m/s to emulate a walking person~\cite{Email2007}. The test experiment is also conducted several rounds per day and repeated in different days. 

The office environment setting includes 5 APs with 3 of them being installed libtins library~\cite{libtins} for crafting the WiFi packets to trigger RTS/CTS process. In contrast, the home environment setting includes 4 APs with 2 of them being able to run RTS/CTS mechanism. Those APs include Alfa AC1200, TPlink AC1750 and Intel 5300 NIC. Fig.~\ref{fig:floor_map}(c), (d) illustrate the heat map of the RSSI collected from a smartphone in both environments, where the signal strength is represented by colour. Because the transmitting power of the phones might vary under different working conditions, some portions of the RSSI heat map are missing or inconsistent. Nevertheless, the signals are consistently stronger in the locations close to AP. The selected APs cover the whole area with a total of 1600~RPs (pink stars) and 900~testing points (solid red line).   

After training, cross validation is carried out on a validation dataset. The validation data consists of a trajectory connecting a set of RPs. The fingerprints of each RP are collected at different time slots from the training dataset, while testing point locations are randomly selected to be different from all RPs in order to approximate our model closer to practical situations. The user location is updated every $\Delta{t}=$1-2~s. Therefore, the sending frequency of crafting RTS frames to the phone is set as $200$ ms per frame to ensure that at least one RSSI scan will occur with every $\Delta{t}$.

\section{Results And Discussions} \label{sec:sim_result}

\subsection{Device Heterogeneity}
\begin{figure}[!t]
\centering
\includegraphics[width=0.8\textwidth]{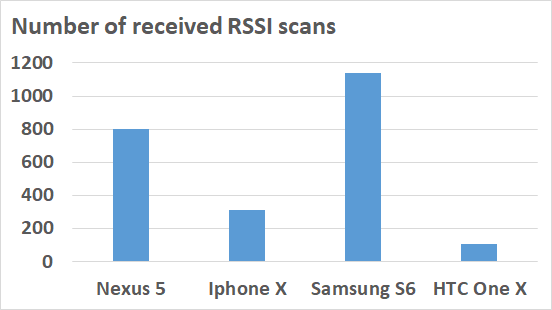}
\caption{Number of received RSSI packets from different phones.}
\label{fig:packets}
\end{figure}

In Subsection \ref{subsec:heter} and \ref{sec:algorithm}, we explained the device heterogeneity problem of RSSI sending from different phones at the same location being significantly different. Additionally, our experiment clearly shows that the number of WiFi packets sent from different phones varies significantly, which also increases estimation error for some phones. Fig.~\ref{fig:packets} illustrates the number of received RSSI packets from different phones in one minute with the same AP at an RTS sending rate of $200$~ms per frame. During the data collection, all the phones are inactive, and the screen is on. Due to the difference of software configuration and operating system of each phone, the number of received packets varies substantially. For example, Samsung S6 has received more than  $1,000$ frames, which is $10$ times more than the frames received by HTC One X. On the other hand, iPhone X has received around $250$ frames, while the number of frames received by Nexus 5 is around $800$. In the latter sections, the experimental results prove that our proposed localization system works well with all these phone models.

\subsection{MAC Randomization}
Starting from Android 8 and above, Android phones have the option to randomize MAC addresses. Currently, the randomly generated MAC address is persistent to the same SSID as long as the phone connects to the same router~\cite{android}.  Therefore, our RTS/CTS mechanism remains valid if the phone WiFi is on and connected to an AP. On the other hand, the MAC address of IOS phones is randomized only when the phone is not associated with an AP~\cite{iphone}.  Since iPhone uses the same phone MAC when it connects to a router, our proposed method is still able to operate.  In the case of IOS phone randomized MAC address when it is not associated with a specific AP, the solution is yet to be found. In the recent reseach~\cite{Brosnan_2022}, the authors proposed a method of clustering algorithm for identifying unique mobile devices from random MAC addresses. It is found that each phone model number has similar set of MAC randomization values. Therefore, both Android and IOS devices can be traced back eventually by collecting enough number of MAC for each model number. However, the drawback is that this method can not work in the case of 2 mobile devices that have the same model number.

\subsection{Antenna Orientation}
\begin{figure}[!t]
     \centering
\subfloat[\label{fig:CSI_angle_am}]
{\includegraphics[width=1\textwidth]{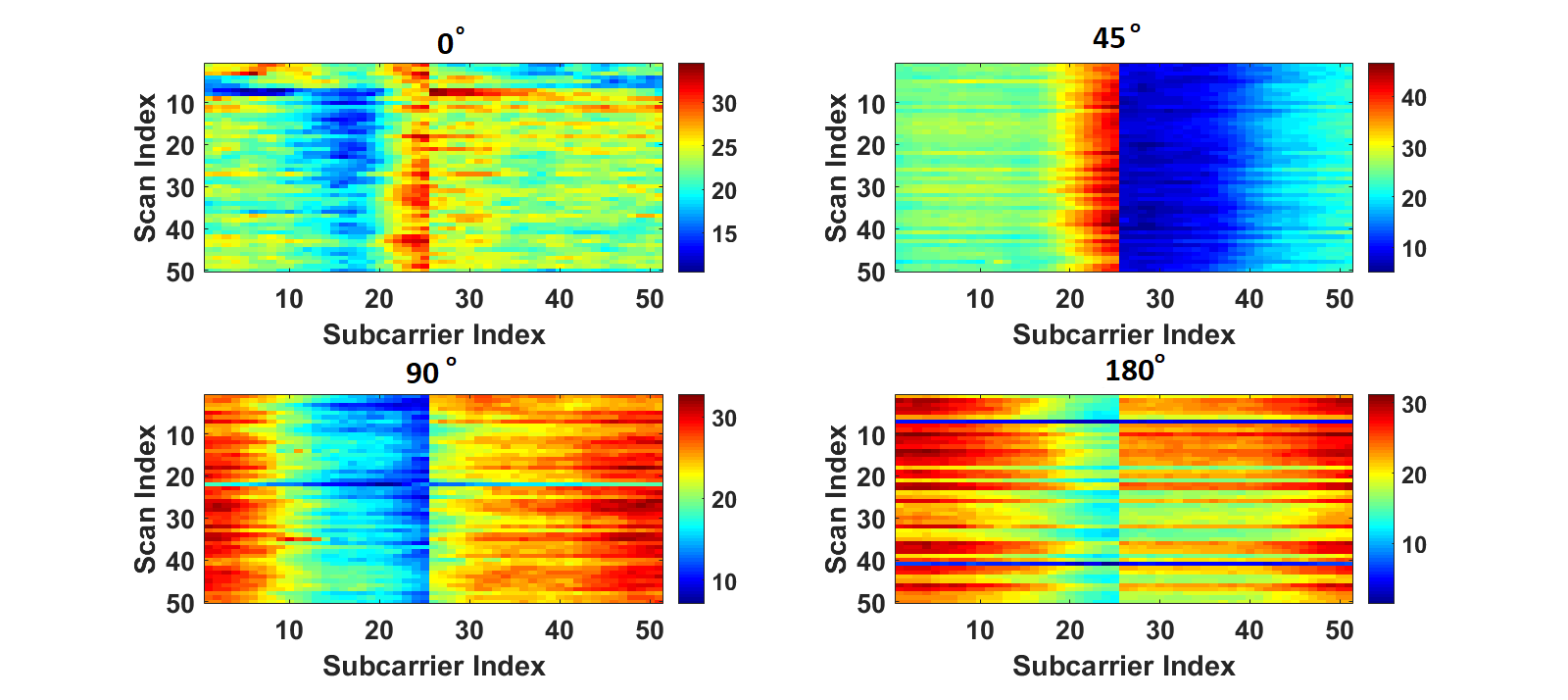}} \quad
\subfloat[\label{fig:CNN_angle_phase}]
{\includegraphics[width=1\textwidth]{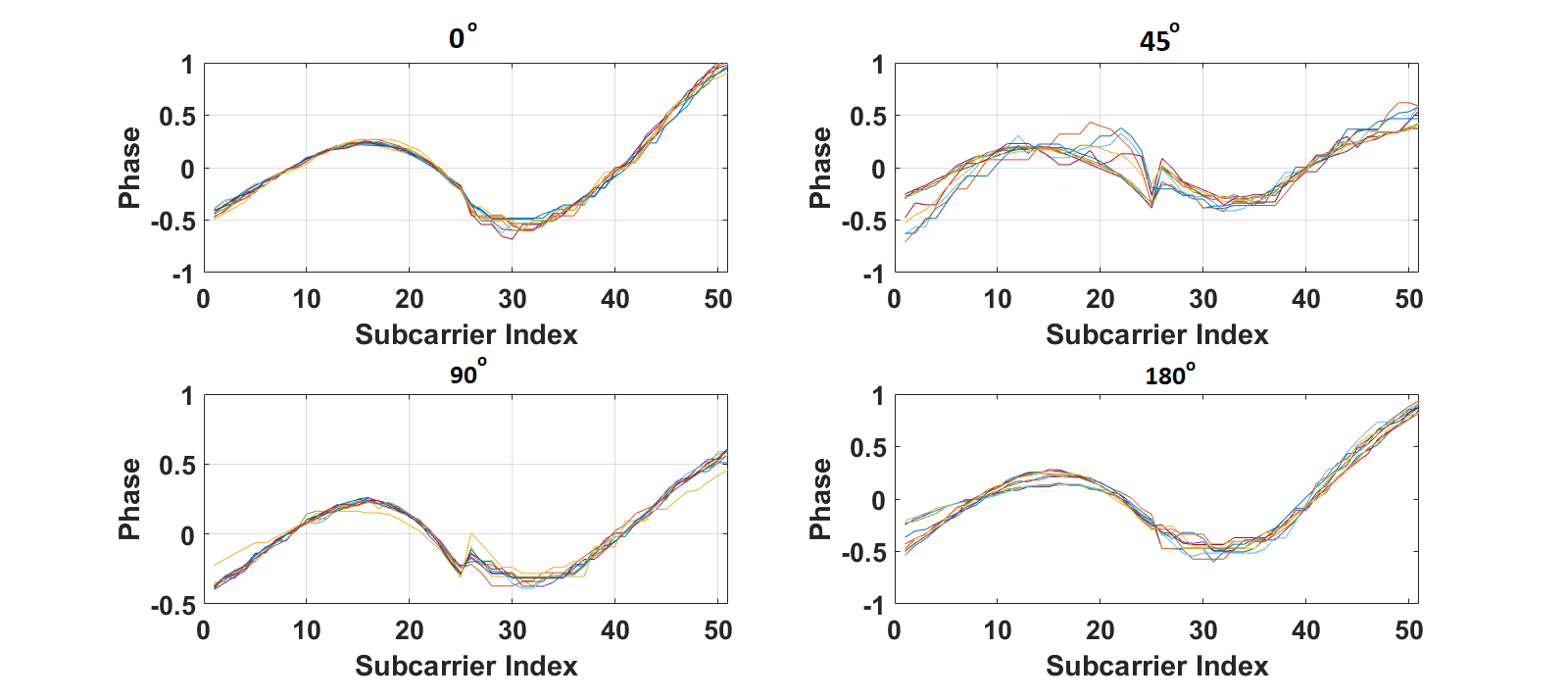}}
\caption{CSI fingerprint features of LG Nexus 5 at a fixed location 1~m away from the router with different orientations (a) CSI amplitude. (b) CSI phase difference.}
     \label{fig:CSI_angle_test}
\end{figure}
The antenna orientation of a mobile device can affect WiFi signal strength significantly~\cite{YogitaChapre2013}.  Fig.~\ref{fig:CSI_angle_test} illustrates the CSI variation of LG Nexus 5 at different orientations angle (0$^\circ$, 45$^\circ$, 90$^\circ$ and 180$^\circ$) to an AP. The distance between the AP and the phone is 1~m and there is a LOS. Fig.~\ref{fig:CSI_angle_test}(a) shows that CSI amplitude vary significantly at different angle. When the angles are 0$^\circ$ and 45$^\circ$, subcarriers around index 25 are among the strongest. However, when the angles are 90$^\circ$ and 180$^\circ$, those become weakest. The correlation of CSI amplitude images between any two of the four angles is only around 30\%. In contrast, Fig.~\ref{fig:CSI_angle_test}(b) illustrates that the CSI phase difference between subcarriers are more stable. The phases of all 4 orientations are similar, with subcarrier 25 at angles 45$^\circ$ and 90$^\circ$ slightly different from the others. The average phase correlations are around 70\% which is more than double compared to those of CSI amplitude. Therefore, CSI phase difference information is far more insensitive to phone orientation than CSI amplitude.

\subsection{Inactive Phones}
\begin{figure}[!t]
\centering
\subfloat[\label{fig:SSP}]
{\includegraphics[width=0.8\textwidth]{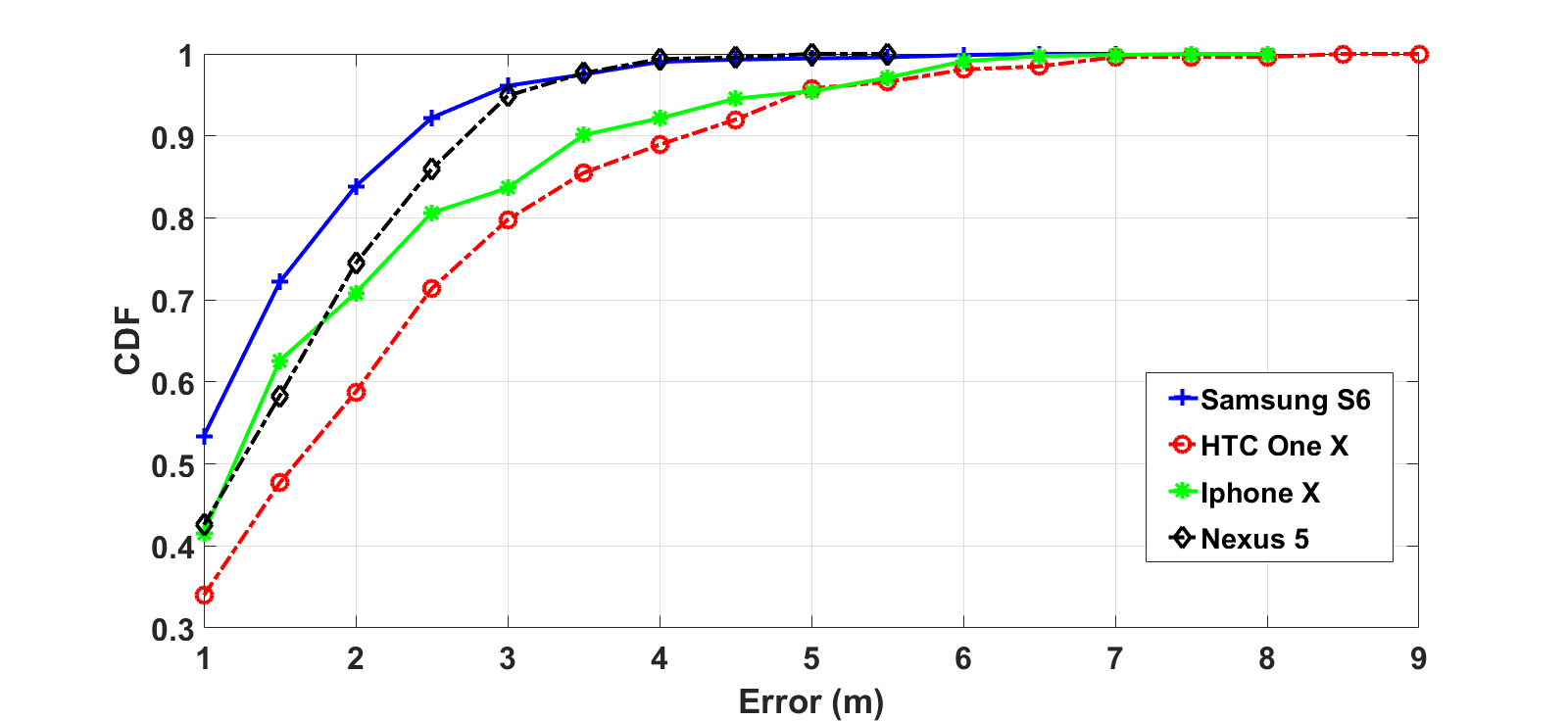}} \quad
\subfloat[\label{fig:LSTM}]
{\includegraphics[width=0.8\textwidth]{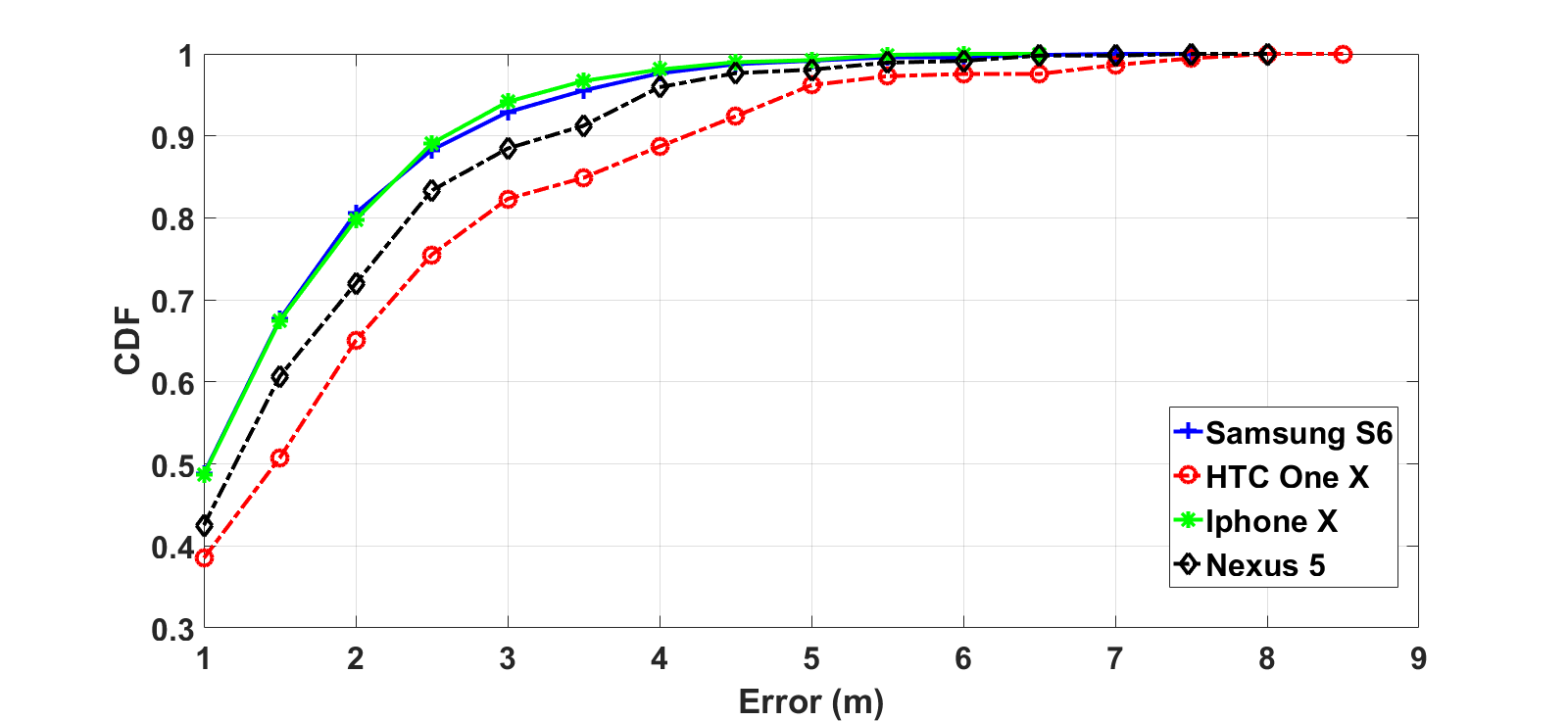}} \quad
\subfloat[\label{fig:SSP_CSI}]
{\includegraphics[width=0.8\textwidth]{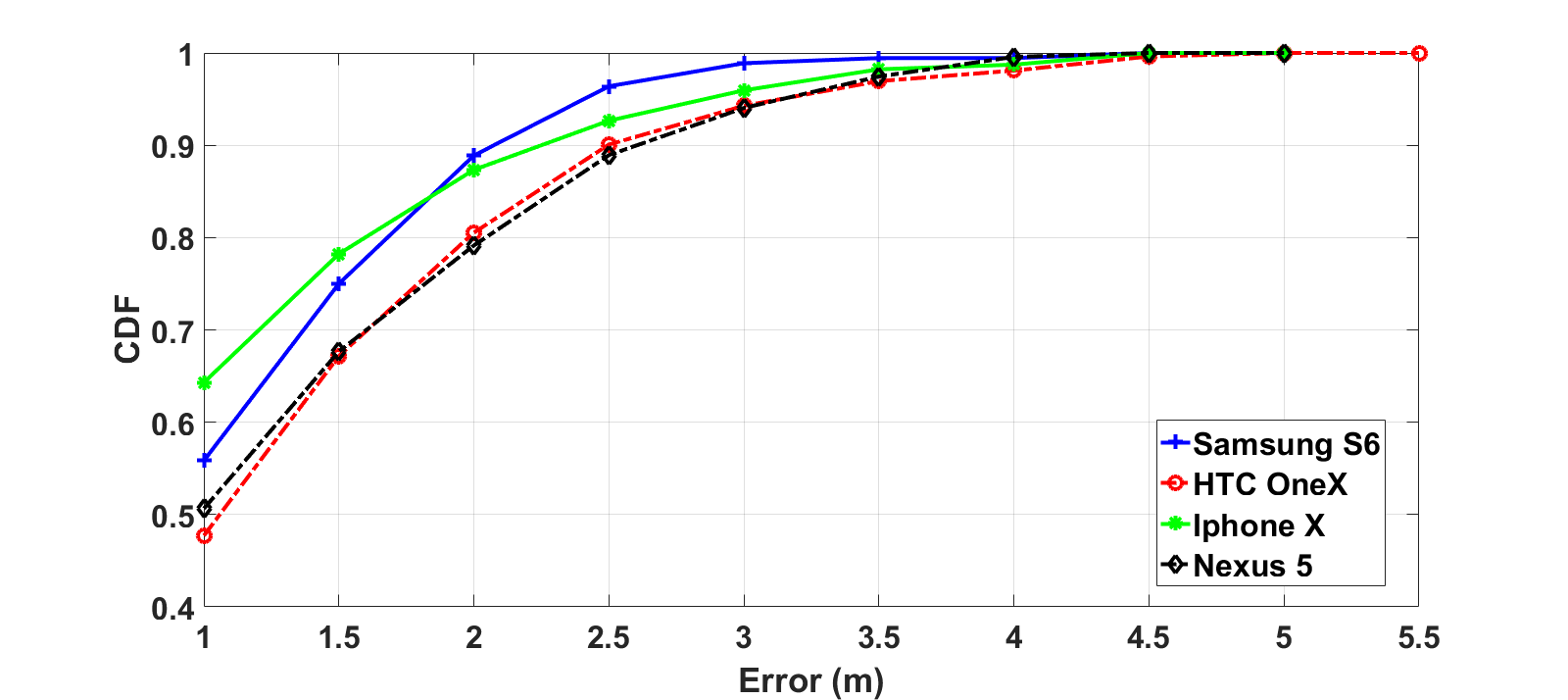}}
\caption{Localization error CDFs of (a) P-MIMO LSTM with different inactive phones. (b) SSP with different inactive phones. (c) Two-step approach with different active phones.}
\end{figure}

When the phone is inactive, we adopt SSP~\cite{Minh2020} and P-MIMO LSTM~\cite{Minh2019} for localization after increasing the data frames through RTS/CTS. For SSP, the Kernel method~\cite{Kushki2007} with Gaussian window and a bandwidth $h=2$ is used since they provide the best accuracy among other combinations according to the experiments in~\cite{Minh2020}. The maximum speed $v_{max}$ the robot can move at is pre-configured to be 4~m/s yielding the maximum distance it can travel during consecutive measurements being $d_{max}=$4~m. Therefore, we use a Gaussian window with the spread $\sigma=d_{max}$. The probability distribution of RSSI is constructed by using the kernel smoothing function in the database of 20,000 scans collected from all phones. Fig.~\ref{fig:SSP} compares the CDF of localization errors of SSP with different phone models. Among all, Samsung S6 provides the best accuracy with 80\% of the localization errors smaller than 2~m, and more than 50\% of testing location errors less than 1~m. On the other hand, Nexus 5 and iPhone X have 80\% less than 2.5~m, while 80\% of HTC One X errors are below 3~m. Besides, the maximum error of HTC One X is 9~m, which is nearly 2 times higher than that of Samsung S6. Table~\ref{table:error2} summarizes the average errors among all phones. As shown, all of them have the average localization error less than 2~m with the best accuracy of 1.2$\pm$0.9~m achieved by Samsung S6.        

\begin{table}
\centering         
\caption{P-MIMO LSTM parameters} \label{table:setup} 
\begin{tabular}{c c}          
\hline
\textbf{Category} & \textbf{Value}  \\ 
RNN type & LSTM \\
Memory length ($T$) & 10 \\
Model & P-MIMO\\
Loss function & RMSE\\ 
Hidden layer (HL)  & 2 \\
Number of neurons for each HL & 100\\
Dropout & 0.2 \\
Optimizer & Adam \\
Learning rate & 0.001 \\
Number of training trajectory & 110,000 \\
Number of validation trajectory & 29,700 \\
$\sigma$ & 4 m \\
$\Delta_t$ & 1 s \\
$d_{max}$ & 4 m \\
 \hline  
\end{tabular} 
\end{table}
The P-MIMO LSTM parameters follow the optimal settings developed in~\cite{Minh2019} and are summarized in Table~\ref{table:setup}. Fig.~\ref{fig:LSTM} compares the CDF of the localization error of the P-MIMO LSTM model. Both Samsung S6 and iPhone X phones performed the best with 80\% localization errors being below 2~m and maximum errors at 6.5~m and 7.5~m respectively. In contrast, HTC has 80\% of errors below 3~m and the maximum error can be as large as 8.5~m.     

\begin{table}[!t]
\centering         
\caption{Average localization errors when the phones are inactive (meter)} 
\label{table:error} 
\begin{tabular}{|l |c |c |c |c|} 
\hline           
\textbf{Method} & \textbf{HTC} & \textbf{Samsung} & \textbf{Nexus 5} & \textbf{iPhone X}\\ 
\hline
SSP & $1.9\pm 1.5$ & $1.2\pm 0.9$ & $1.4\pm 0.9$ & $1.6\pm1.4$\\
P-MIMO LSTM &  $1.8\pm 1.5$ & $1.2\pm 1.0$ & $1.4\pm 1.2$ & $1.2\pm 0.9$\\
\hline         
\end{tabular} 
\end{table}

Table~\ref{table:error} further compares the average errors of different phones using SSP and P-MIMO LSTM.  As shown, in general, P-MIMO LSTM has slightly better performance than the SSP model while the training of SSP is less computational intensive.  In the experiment, the training of P-MIMO LSTM is executed on a computer with an AMD FX-8120 Eight-Core CPU and an Nvidia GTX 1050 GPU. The running time is approximately 4~s per epoch. Therefore, the training time is approximately $\cong$ 1 hour and 6 minutes. Comparing with SSP in training phase, all the kernel density calculation to estimate RSSI PDF is finished less than 5 minutes for all $1,600$ RPs.   

\subsection{Active Phones} \label{sec:trans_result}
\begin{table}[!t]
\centering         
\caption{Localization Errors Comparison Between Cases (meter)} 
\label{table:error2} 
\begin{tabular}{l c c c c} 
\hline           
\textbf{Method} & \textbf{HTC} & \textbf{Samsung} & \textbf{Nexus 5} & \textbf{iPhone X}\\ 
SSP (Inactive phone) & $1.9\pm 1.5$ & $1.2\pm 0.9$ & $1.4\pm 0.9$ & $1.6\pm1.4$\\
SSP (Active phone) &  $1.2 \pm 0.9$  & $1.0 \pm 0.7$  & $1.2 \pm 0.9 $ & $1.0 \pm 0.8$ \\
\hline         
\end{tabular} 
\end{table}

As explained in Subsection \ref{sec:trans}, localization for active phones takes a two-step approach. During the first step, RSSI is selected as fingerprints and SSP is used to select $K=5$ most probable locations. During the second step, the CSI amplitude and phase of the testing point is correlated with the CSI information of the $K$ neighbour locations using Pearson coefficient~\cite{Minh2018} and the one that provides the largest coefficient is predicted to be the location. Table~\ref{table:error2} illustrates the average error comparisons between active phone localization and inactive phone localization. With the additional information from CSI for active phone localization, the accuracy enhances approximately 20\%-40\%. Among all phones, HTC and iPhone X have the largest improvement with the average error decreasing from 1.9$\pm$1.5~m and 1.6$\pm$1.4~m for the inactive phones to 1.2$\pm$0.9~m and 1.0$\pm$0.8~m for the active phones. On the other hand, the localization accuracy of Samsung and Nexus 5 enhance slightly by 0.2~m.         

Fig.~\ref{fig:SSP_CSI} illustrates the CDF errors of 4 phones. For active phone localization, the performance of all phones are similar to each other in contrast to the inactive phone localization. This is caused by the active phones transmitting a similar number of data frames. Additionally, CSI data provides important information to make the localization more accurate. Overall, 80\% average errors of HTC and Nexus 5 are below 2.5~m, while the ones of Samsung and iPhone X are below 1.8~m. The maximum errors of HTC are 5.5~m, whereas those of the other phones are around 5~m. 

\section{Conclusion}
In conclusion, we have developed a comprehensive practical passive localization.  The localization accuracy has been investigated through extensive on-site experiments using autonomous robots with multiple phones in hundreds of testing locations. With the introduction of the RTS/CTS mechanism, continuous localization of a mobile user is achieved with an accuracy of 1.5~m even when the phone is inactive. For active phones, the 2-step approach that incorporates the CSI as part of the WiFi fingerprints further enhances the accuracy to 0.8~m. In the future work, more investigation will be conducted in the case of randomized MAC address phone that does not associate with a specific AP. On the other hand, more experiments will be conducted to research about the effects of different environmental conditions and weather conditions on the accuracy of our indoor localization method.       
\section*{Funding}
 This work was supported in part by the Natural Sciences and Engineering Research Council of Canada under Grant 520198, Fortinet Research under Contract 05484 and NVidia under GPU Grant program.

\section*{Declarations}

\begin{itemize}
\item This work was supported in part by the Natural Sciences and Engineering Research Council of Canada under Grant 520198, Fortinet Research under Contract 05484 and NVidia under GPU Grant program.
\item The authors declare no competing interests.
\item Ethics approval and consent to participate
\item The dataset generated and analyzed during the current study are available from the corresponding author on reasonable request.
\item The code is available upon request.

\item Author contribution: M.H. and B.Y. conducted the research, M.H. wrote the main manuscript text. X.D. and T.L. supervised the project and revised the manuscript with M.H. and B.Y.. M.H., B.Y., K.R., A.E., X.D., T.L., R.W., and K.T. analyzed the data, reviewed the manuscript.
\end{itemize}

\end{document}